\documentclass[prd,twocolumn,nofootinbib,superscriptaddress,amsmath,amssymb]{revtex4-1}

\usepackage{mathtools}
\usepackage{graphics}
\usepackage{graphicx}
\graphicspath{{./images/}}
\usepackage{dcolumn}
\usepackage{bm}
\usepackage{dsfont} 
\usepackage{amsmath,amssymb}
\usepackage{hyperref}
\usepackage{tabularx}
\usepackage{epstopdf}
\usepackage[normalem]{ulem}
\usepackage[usenames]{color}
\usepackage{multirow}
\usepackage{makecell}
\usepackage{diagbox}
\usepackage[abs]{overpic}
\allowdisplaybreaks
\hypersetup{
    colorlinks=true,
    linkcolor=blue,
    filecolor=magenta,      
    urlcolor=blue,
    citecolor=blue
}
\definecolor{red(ncs)}{rgb}{0.77, 0.01, 0.2}

\usepackage{array}
\newcolumntype{C}[1]{>{\centering\let\newline\\\arraybackslash\hspace{0pt}}m{#1}}

\newcommand{\hh}{\mathbf{h}}
\newcommand{\dd}{\mathbf{d}}
\newcolumntype{C}[1]{>{\centering\arraybackslash}m{#1}}

\def\emri#1{Extreme Mass-Ratio Inspiral#1 (EMRI#1)\gdef\emri{EMRI}}
\def\imbh#1{Intermediate Mass Black Hole#1 (IMBH#1)\gdef\imbh{IMBH}}
\def\smbh#1{supermassive black hole#1(SMBH#1)\gdef\smbh{SMBH}}
\def\bbh#1{binary black hole#1 (BBH#1)\gdef\bbh{BBH}}
\def\imbhb#1{intermediate mass black hole binary#1 (IMBHB#1)\gdef\imbhb{IMBHB}}
\def\hmns#1{hypermassive neutron star#1 (HMNS#1)\gdef\hmns{HMNS}}
\def\bh#1{black hole#1 (BH#1)\gdef\bh{BH}}
\def\ns#1{neutron star#1 (NS#1)\gdef\ns{NS}}
\def\hmns#1{hyper-massive neutron star#1 (HMNS#1)\gdef\hmns{HMNS}}
\def\bhns#1{black hole-neutron star#1 (BHNS#1)\gdef\bhns{BHNS}}
\def\nsbh#1{neutron star-black hole#1 (NSBH#1)\gdef\bhns{NSBH}}
\def\bns#1{binary neutron star#1 (BNS#1)\gdef\bns{BNS}}
\def\gw#1{gravitational wave#1 (GW#1)\gdef\gw{GW}}
\def\pnw#1{post-Newtonian#1 (PN#1)\gdef\pnw{PN}}
\def\eos#1{equation of state#1 (EOS#1)\gdef\eos{EOS}}
\def\gpu#1{graphics processing unit#1 (GPU#1)\gdef\gpu{GPU}}
\def\gr#1{General Relativity#1 (GR#1)\gdef\gr{GR}}
\def\cbc#1{compact binary coalescence#1 (CBC#1)\gdef\cbc{CBC}}
\def\nr#1{Numerical Relativity#1 (NR#1)\gdef\nr{NR}}
\def\hom#1{Higher Order Mode#1 (HOM#1)\gdef\hom{HOM}}
\def\qnm#1{Quasi Normal Mode#1 (QNM#1)\gdef\qnm{QNM}}
\def\emw#1{Electromagnetic Waves#1 (EMW#1)\gdef\emw{EMW}}
\def\snr#1{Signal to Noise Ratio#1 (SNR#1)\gdef\snr{SNR}}
\def\gr#1{General Relativity#1 (GR#1)\gdef\gr{GR}}
\def\psd#1{Power Spectral Density#1 (PSD#1)\gdef\psd{PSD}}
\def\asd#1{Amplitude Spectral Density#1 (ASD#1)\gdef\asd{ASD}}
\def\grb#1{gamma-ray burst#1 (GRB#1)\gdef\grb{GRB}}
\def\far#1{False Alarm Rate#1 (FAR#1)\gdef\far{FAR}}
\def\pe#1{Parameter Estimation#1 (PE#1)\gdef\pe{PE}}
\def\mcmc#1{Makov Chain Monte Carlo#1 (MCMC#1)\gdef\mcmc{MCMC}}
\def\mlw#1{maximum likelihood {\tt LALInference} waveform#1 (MLW#1)\gdef\mlw{MLW}}
\def\mbw#1{median {\tt BayesWave} waveform#1 (MBW#1)\gdef\mbw{MBW}}

\newcommand{\CCA}{\affiliation{Center for Computational Astrophysics, Flatiron Institute, 162 5th Ave, New York, NY 10010}}

\newcommand{\GA}{\affiliation{Center for Relativistic Astrophysics, Georgia Institute of Technology, Atlanta, Georgia 30332, USA}}

\newcommand{\OzGrav}{\affiliation{OzGrav, School of Physics, University of Melbourne, Parkville, Victoria 3010, Australia}}

\newcommand{\NASA}{\affiliation{NASA Marshall Space Flight Center, Huntsville, AL 35812, USA}}

\newcommand{\MSU}{\affiliation{eXtreme Gravity Institute, Department of Physics, Montana State University, Bozeman, Montana 59717, USA}}

\begin{document}

\title{Reconstructing gravitational wave signals from binary black hole mergers with minimal assumptions}

\author{Sudarshan Ghonge}
\GA

\author{Katerina Chatziioannou}
\CCA

\author{James A. Clark}
\GA

\author{Tyson Littenberg}
\NASA

\author{Margaret Millhouse}
\OzGrav

\author{Laura Cadonati}
\GA

\author{Neil Cornish}
\MSU

\date{\today}


\begin{abstract}
We present a systematic comparison of the \bbh{} signal waveform reconstructed by two independent and complementary approaches used in LIGO and Virgo source inference: a template-based analysis, and a morphology-independent analysis. 
We apply the two approaches to real events and to two sets of simulated observations made by adding simulated \bbh{} signals to LIGO and Virgo detector noise. 
The first set is representative of the 10 \bbh{} events in the first Gravitational Wave Transient Catalog (GWTC-1).   
The second set is constructed from a population of \bbh{} systems with total mass and signal strength in the ranges that ground based detectors are typically sensitive. 
We find that the reconstruction quality of the GWTC-1 events is consistent with the results of both sets of simulated signals. 
We also demonstrate a simulated case where the presence of a mismodelled effect in the observed signal, namely higher order modes, can be identified through the morphology-independent analysis.
This study is relevant for currently progressing and future observational runs by LIGO and Virgo.
\end{abstract}

\maketitle


\section{Introduction}
\label{sec:intro}

The first Gravitational Wave Transient Catalog (GWTC-1) \cite{LIGOScientific:2018mvr} published by the LIGO and Virgo Collaboration \cite{TheLIGOScientific:2014jea,TheVirgo:2014hva} includes signals from ten Binary Black Hole (\bbh{}) sources and one \bns{} system. Along with \nsbh{} binaries, these sources are referred to as \cbc{s}. Ground based detectors are most sensitive to transient signals from \cbc{} systems with total mass $M_{T}$ in the stellar mass range ($10 M_{\odot} \lesssim M_{T} \lesssim 100 M_{\odot}$). The \gw{} emission becomes loudest in the sensitive frequency band ($20 - 1000$ Hz) \cite{Harry_2010} milliseconds to minutes before the merger, just as the \gw{} emission reaches peak amplitude.

There are two main types of transient \gw{} analysis: targeted template-based matched-filter ``\cbc{}'' analyses which use physically-motivated waveform models \cite{Allen:2005fk}, and morphology-independent ``burst'' analyses \cite{Anderson:2000yy,Andersson:2013mrx}.  The models used in \cbc{} analyses \cite{Usman:2015kfa,Messick:2016aqy} are semi-analytical solutions of \gr{} that combine aspects of analytical post-Newtonian theory to model the inspiral, and \nr{} \cite{Pretorius:2005gq, Campanelli:2005dd, Baker:2005vv} to capture the highly non-linear late inspiral and merger phases \cite{Bohe:2016gbl}. The \cbc{} templates account for the dominant $(l,|m|)=(2,2)$ mode in the spherical harmonic formulation of \gw{} radiation.  Burst analyses model \gw{s} as a superposition of a number of suitable basis functions parameterized by observable quantities such as amplitude and frequency \cite{Klimenko_2008, Lynch:2015yin}.  The inexact match of the basis functions with underlying \gw{} signals results in generally lower intrinsic sensitivity than targeted \cbc{} searches but the larger number of degrees of freedom allows for the recovery of un-modelled waveform phenomenology and, potentially, new physics.  Burst methods are also used to search for \gw{} signals from sources such as supernovae~\cite{New:2002ew} and the post-merger phase of binary neutron star coalescence, where the physics is too uncertain to develop a sufficiently robust matched-filter template \cite{Abbott:2018wiz, Abbott:2017dke, Chatziioannou:2017ixj, Torres-Rivas:2018svp}.

Following the detection of a \gw{} signal in the data, \pe{} analysis is performed by {\tt LALInference} \cite{Veitch:2014wba}, which uses \cbc{} models to sample the posterior probability distribution (PDF) of the physical parameters, e.g. masses and spins, using stochastic samplers such as \mcmc{}\cite{gilks1995markov, Metropolis:1953am, geman1987stochastic,Rover:2006ni, Christensen:2001cr,Rover:2006bb, Raymond:2009cv} and Nested Sampling \cite{skilling2006nested, Veitch:2009hd}. The resulting PDF is used in studies including formation scenarios, rates and tests of \gr{}. 

A ``Burst" PE analysis is performed by {\tt BayesWave} \cite{cornish2015bayeswave, Littenberg:2014oda}. {\tt BayesWave} models the signal waveform as a sum of Morlet Gabor wavelets \cite{morlet1982wave} and uses a trans-dimensional Reversible Jump Markov Chain Monte Carlo (RJMCMC) to sample the parameters as well as the number of the wavelets \cite{green1995reversible}. {\tt BayesWave} will reconstruct any feature in the data that is coherent across the detector network if the feature is loud enough compared to the background noise. This makes the wavelet model flexible enough to fit a wide range of signal morphologies. For the case of \bbh{} signals, it is most sensitive to times close to the merger where the amplitude peaks. 

\begin{figure*}[]
    \centering
    \includegraphics[width=2\columnwidth,clip=true]{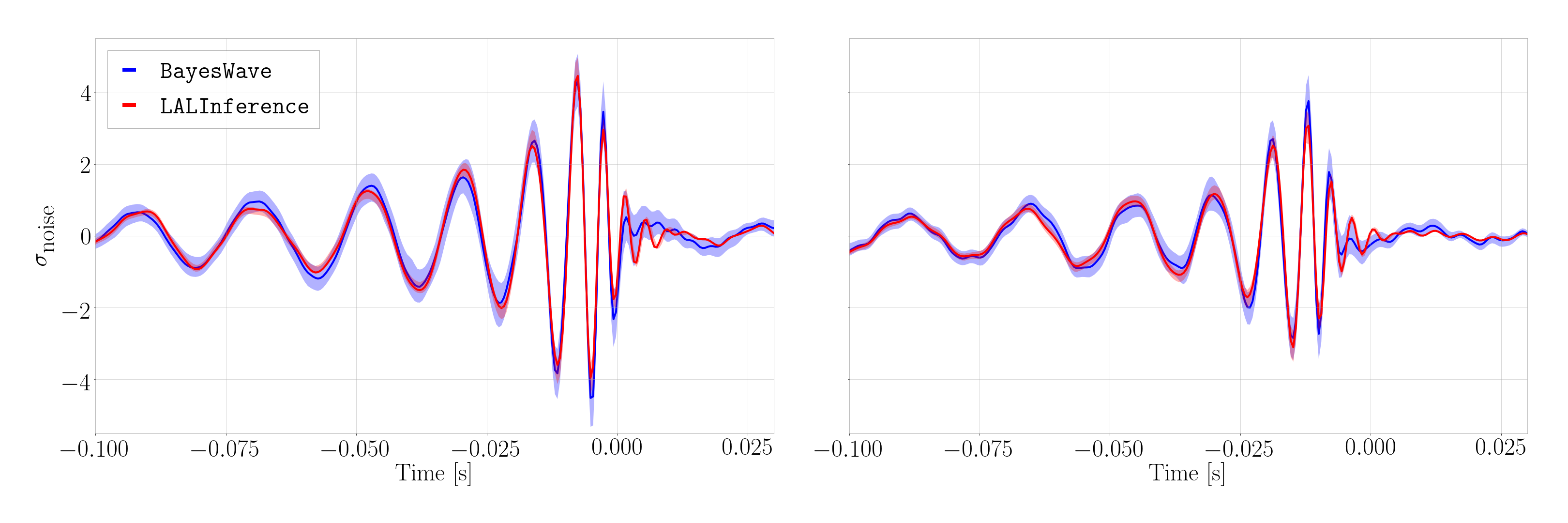}
    \caption{{\tt LAInference} {\tt IMRPhenomPv2} (red) \cite{Hannam:2013oca} and {\tt BayesWave} (blue) reconstructions of GW150914 \cite{Abbott:2016blz} using publicly released data from GWTC-1 \cite{gwtc1:fig10}. The left and right panels respectively show the waveform in LIGO Hanford and LIGO Livingston. The x-axis represents the time in seconds before the coalescence. The y axis represents the strain amplitude whitened using a filter which is the inverse amplitude spectral density (ASD). The units are in multiples of the standard deviation of the noise. }
    \label{fig:gw150914_reconstruction}
\end{figure*}

Wavelets and \cbc{} waveforms provide complementary means to study \gw{} signals. Fig. \ref{fig:gw150914_reconstruction} shows the waveform reconstructions and their 90\% credible intervals given by {\tt LALInference} using the precessing, dominant mode approximant {\tt IMRPhenomPv2} \cite{Hannam:2013oca} from the Phenom waveform family, and by {\tt BayesWave} for GW150914 \cite{Abbott:2016blz}. Waveform reconstruction plots that similarly illustrate the agreement between \cbc{} and burst reconstructions and have been used in works on GW150914 \cite{TheLIGOScientific:2016uux}, GW170104  \cite{Abbott:2017vtc}, GW170814 \cite{Abbott:2017oio}, GW170729 \cite{Chatziioannou:2019dsz}, and GWTC-1 \cite{LIGOScientific:2018mvr}. Comparing wavelet-based and waveform-based signal reconstructions serves as a consistency check for the signal morphology. A general feature of these plots is that the reconstructions agree at times close to the merger where the signal is strong, but do not necessarily have to agree where the signal is weak. This is because accuracy of {\tt BayesWave} of reconstruction a feature depends on the loudness of the feature. We also note that the {\tt BayesWave} credible intervals are broader than the {\tt LALInference} credible intervals since the former allows for more flexibility in the waveform morphology.

Reference \cite{TheLIGOScientific:2016uux} studies the agreement between a set of simulated \gw{} signals injected into real data and the reconstructions obtained using {\tt BayesWave}. A related test of signal consistency is the residuals test which uses {\tt BayesWave} to analyze the residual obtained by subtracting from the detector data the reconstructed \cbc{} signal. The result is then compared to the same analysis on surrounding noise to quantify the evidence for any residual excess.  This test has been employed in \cite{TheLIGOScientific:2016src, LIGOScientific:2019fpa, Abbott:2017vtc}.

This paper presents a systematic performance comparison of the two algorithms applied to \bbh{} systems. It provides the context in which the reconstructions of future gravitational wave events can be evaluated, which is particularly timely given the approximately weekly \bbh{} detections during the third LIGO and Virgo observing run. Instead of qualitative plot comparisons, we use a quantitative comparison metric that is the \textit{overlap}, which is the noise weighted inner product of waveforms reconstructed by each algorithm. Simulated \bbh{} \gw{} signals are added to detector noise from the LIGO and Virgo detectors.  These ``injections'' are then analyzed using {\tt LALInference} and {\tt BayesWave}. We perform two types of injections: in the first, we inject populations of signals whose physical parameters are drawn from the posterior probability distributions inferred from GWTC-1 events \cite{Vallisneri_2015}.  We also analyze a population of \bbh{} injections whose masses are drawn uniformly from ranges which explore ground based detector sensitivities and signal durations. 

We find that the waveform reconstructions of events in the GWTC-1 catalog are consistent, within 90\% credibility,  with expectations based on our simulations of similar signals.  Analysis of signals drawn from across the mass spectrum also illustrates that {\tt BayesWave} performs significantly better for higher mass systems while the template-based {\tt LALInference} reconstructions are relatively insensitive to the mass ranges explored in this study. This is to be expected, since shorter-duration with fewer cycles most closely resemble the wavelet basis used by {\tt BayesWave}, bringing the analysis closer to matched-filtering.

There remain known physical effects, such as precession, orbital eccentricity, extreme mass ratios which have historically been difficult to incorporate into analytical models for \bbh{} \gw{} waveforms.  Less certain effects, such as deviations from \gr{}, are still more difficult to model.  With developments in technology such as LIGO A+ \cite{barsotti2018a+}, and third generation detectors such as the Einstein Telescope and Voyager \cite{punturo2010einstein, collaboration2016instrument}, the network of ground based detectors will reach sensitivities where these effects will in principle, be loud enough to cause significant disagreement between the reconstrucions given by the \cbc{} and model-independent analyses.  As an illustration of such a scenario, we analyze a numerical relativity waveform from the Georgia Tech catalog\cite{Jani:2016wkt}. In this system, \hom{s} contribute a substantial fraction of the total signal-to-noise ratio \cite{Bustillo:2016gid}.  While there now exist waveform templates which accurately model \hom{s} \cite{London:2017bcn, Cotesta:2018fcv}, analysis of this signal with a more rudimentary waveform model \cite{Hannam:2013oca} is a convenient way to highlight what a disparity in {\tt LALInference} and {\tt BayesWave} reconstructions would look like. We analyze the performance of {\tt LALInference} and {\tt BayesWave} when this waveform is injected into data and find that the latter is able to reconstruct the waveform more accurately due to its flexibility.

Section \ref{sec:methodology} delves into the details of {\tt LALInference} and {\tt BayesWave}, their waveform models, sampling techniques, and calculation of the overlap. Section \ref{sec:injections} describes the set of injections in detail. Section \ref{sec:results} discusses the results and inferences. Section \ref{sec:disagreement} briefly discusses the performace comparison of the two algorithms when \hom{s} are included in the injection. Section \ref{sec:conclusion} concludes the paper and discusses possible future work.


\section{Methodology}

\label{sec:methodology}
The properties of a detected signal are inferred by modeling the detector data $\dd$ with the parameterized waveform $\hh(\theta)$. The boldface here is to emphasize that $\dd$ and $\hh$ represent quantities in multiple detectors. Here $\theta = \{\theta_1, \theta_2, ...,\theta_N\}$ represents a point the parameter space of the underlying \cbc{} system in the case of {\tt LALInference} such as masses and spins, or the parameter space of wavelets in the case of {\tt BayesWave} such as the central frequency, amplitude and number of wavelets. The data $\dd$ are assumed to be a time series that contains the true \gw{} signal, plus additive stationary Gaussian noise characterized by the one-sided noise \psd{} $S_n(f)$. We are interested in sampling the posterior probability distribution function of $\hh$ given $\dd$. Accoriding to Bayes' theorem \cite{bayes1763lii, jaynes1996probability}:
\begin{equation}
    p(\hh | \dd) = \frac{p(\hh) p(\dd | \dd)}{p(\dd)},
    \label{eq:bayes_thm}
\end{equation}
where $p(\hh)$ is the prior knowledge about the system. $p(\dd |\hh)$ is the likelihood function, the probability of obtaining data $\dd$ given the signal $\hh$: 

\begin{equation}
    p(\dd|\hh) \propto \exp\left( - \frac{1}{2} \langle \dd - \hh |\dd - \hh \rangle \right),
    \label{eq:likelihoood}
\end{equation}
where $\langle \cdot | \cdot \rangle$ on quantities with boldface indicates the noise weighed inner product over the network of detectors given by:
\begin{equation}
    \langle \mathbf{a}|\mathbf{b} \rangle =  \sum_{i}^{n} \langle a^i | b^i \rangle, 
    \label{eqn:overlap}
\end{equation}

here $i$ sums over all $n$ detectors in the network, and $\langle a^i|b^i \rangle$ is the inner product in an individual detector defined in as
\begin{equation}
\langle a^i|b^i \rangle \equiv 4 \Re \int_{0}^{\infty} \frac{\tilde{a}^i(f)\tilde{b}^{i*}(f)}{S_n^i(f)}df,
\end{equation}

$\tilde{a}^{i}(f)$ is the Fourier transform of time series $a^{i}$, and the superscript $*$ denotes the complex conjugate. $S_n^i(f)$ is the \psd{} of the $i^{\textrm{th}}$ detector. Dividing by the \psd{} effectively reweights the integral towards frequencies  where the detectors are most sensitive. The optimal \snr{} is defined as
\begin{equation}
    \rho = \sqrt{\langle \dd| \dd \rangle},
\end{equation}
and is often used as a figure of merit for the strength of the signal in the detector.

The signal in the $i^{th}$ detector, $h^{i}$, is obtained by projecting the ``plus" ($h_{+}$) and ``cross" ($h_{\times}$) using the sky-location dependent antenna pattern functions $F_{+}^{i}$ and $F_{\times}^{i}$:
\begin{equation}
h^{i} = F^{i}_{+}(\theta, \phi, \psi)\, h^{i}_{+} + F^{i}_{\times}(\theta, \phi, \psi)\, h^{i}_{\times}
\end{equation}

The computational cost of estimating the likelihood function using deterministic methods is high, as the number of valuations required to explore the parameter space on a fixed grid grows exponentially with the number of dimensions. This can become prohibitively expensive beyond a few dimensions. Therefore, sampling-based methods such as Markov Chain Monte Carlo (MCMC) \cite{gilks1995markov,Metropolis:1953am,geman1987stochastic,Rover:2006ni, Christensen:2001cr,Rover:2006bb, Raymond:2009cv} and Nested Sampling \cite{skilling2006nested, Veitch:2009hd} are often used.

\subsection{{\tt LALInference}}
\label{sec:LI}

{\tt LALInference} \cite{Veitch:2014wba} models the  signal $\dd$ in the detector data as a \cbc{} \gw{} signal described by GR. It uses analytical or semi-analytical approximants to construct the signal waveform. To sample the parameter space it uses two main techniques: Nested Sampling \cite{skilling2006nested, Veitch:2009hd} and MCMC \cite{gilks1995markov,Metropolis:1953am,geman1987stochastic,Rover:2006ni, Christensen:2001cr,Rover:2006bb, Raymond:2009cv}. 

The parameter samples of GWTC-1 use the precessing, dominant mode approximants from the two main families: {\tt IMRPhenomPv2} \cite{Hannam:2013oca} from the Phenom family,  and {\tt SEOBNRv3} \cite{Taracchini:2013rva, Pan:2013rra} from the EOB-NR family. For this paper we use the {\tt IMRPhenomPv2} samples to perform injections, and use a reduced order quadrature (ROQ) \cite{Smith:2016qas} of {\tt IMRPhenomPv2} to compute the likelihood while analyzing with {\tt LALInference}. The ROQ  reduces the computational cost of parameter estimation by reducing redundant computations. We do not use the {\tt SEOBNRv3} approximant for recovery as it is computationally more expensive. Studies such as \cite{LIGOScientific:2018mvr} have shown that {\tt IMRPhenomPv2} and {\tt SEOBNRv3} samples for \bbh{} systems in GWTC-1 broadly agree with each other.

\subsection{{\tt BayesWave}}
\label{sec:bw}
{\tt BayesWave} \cite{cornish2015bayeswave, Littenberg:2014oda} models the signal $\dd$ in the detectors as a summation of Morlet Gabor wavelets, the number and parameters of which are marginalized over using the reversible jump markov chain monte carlo (RJMCMC) sampler. 

The signal model consists of a variable number of wavelets, where each wavelet, $\Psi$, is described by five parameters: the central time $t_0$, the central frequency $f_0$, the quality factor $Q$, the amplitude $A$, and the phase offset $\phi_0$. In the frequency domain, the wavelet is given by

\begin{multline}
\widetilde{\Psi}(f;A,Q,f_0, t_0, \phi_0) = \\
\frac{\sqrt{\pi} A \tau}{2}  e^{-\pi^2 \tau^2 (f-f_0)^2} e^{i(\phi_0 +  2 \pi (f - f_0) t_0) },
\label{eq:bw_wavelet}
\end{multline}
where $\tau = Q/2 \pi f_0$ and and $\tilde{\cdot}$ represents the frequency domain version of any quantity. Assuming an elliptically polarized \gw{} signal, the plus component ($h_{+}$) of the \gw{} strain is given by $\widetilde{h}_{+} = \sum_{j=0}^{N} \Psi_{j}$, where $N$ is the number of wavelets that describe the signal model. The cross component ($h_{\times}$) is given by $\widetilde{h}_{\times} = i e \widetilde{h}_{+}$, where $e$ is the ellipticity paramter which is also sampled over. Details of the wavelet model used in {\tt BayesWave} can be found in \cite{cornish2015bayeswave}.
A generalization of the wavelet model is the chirplet model which includes a time dependent frequency component \cite{Millhouse:2018dgi}. 

Since we are testing the infrastructure as employed by past LIGO and Virgo Collaboration papers, we limit our analysis to the wavelet model. Past studies such as \cite{Chatziioannou:2019dsz} have shown that the wavelet and chirplet models have similar levels of agreement with \cbc{} waveforms for the observed \bbh{} systems. We also limit ourselves to the frequency independent ellipticity ($e$) assumption. {\tt BayesWave} was initially developed using the elliptical polarization assumption since the early era of \gw{} astronomy had only two nearly-aligned LIGO detectors which resulted in poor polarization sensitivity. Recent works such as \cite{LIGOScientific:2020stg} show that \hom{s} are measurable with the current detector network sensitivities and it is important to relax the ellipticity constraint, where the parameters of $\widetilde{h}_{+}$ and $\widetilde{h}_{\times}$ are independently sampled. At the time of preparing this work, development towards this independent polarization model is complete and has been demonstrated to work. It will be discussed in future works.

\subsection{Overlap}
To quantify the agreement between {\tt LALInference} and {\tt BayesWave}, we use point estimates of the signal waveform from each. In the case of {\tt LALInference}, we use the posterior sample for which the likelihood function described in Eq \ref{eq:likelihoood} is maximum, which we will call the \mlw{}. We caution that this is a good approximation of, but not necessarily, the true maximum, as {\tt LALInference} is a posterior distribution inferring algorithm, rather than a peak finding algorithm. For {\tt BayesWave} we use the estimate obtained by taking the median of the waveform value at every time index from the whole set of samples. We call this the \mbw{}. We do not use the maximum likelihood {\tt BayesWave} waveform since unlike \cbc{} waveforms, the wavelets are ``nuisance parameters" that do not have any physical meaning themselves. Instead it is the fit waveform that is fundamentally of interest. The \mbw{} is a collective estimate across samples that is stable because it is relatively immune to the stochastic fluctuations of the variable dimensional sampler. 

We quantify the agreement between the \mlw{} ($\hh_\textrm{LI}$) and the \mbw{} ($\hh_\textrm{BW}$) by computing the \textit{overlap} over the network of detectors \cite{Apostolatos:1995pj}.
\begin{equation}
{\cal{O}}_\textrm{B,L} \equiv \frac{\langle \hh_\textrm{LI} | \hh_\textrm{BW} \rangle}{\sqrt{\langle \hh_\textrm{LI}|\hh_\textrm{LI} \rangle \langle \hh_\textrm{BW}|\hh_\textrm{BW} \rangle}},
\label{eq:network_overlap}
\end{equation}

We use the parameterized version given by the {\tt BayesLine} \cite{Littenberg:2014oda} algorithm which is a fully integrated in {\tt BayesWave}. {\tt BayesLine} models the \psd{} with two components: a cubic spline to fit the broad band noise, and a sum of Lorentzians to fit the narrow band spectral lines. The number and location of Lorentzians and cubic spline control plots are again determined with a RJMCMC. This \psd{} estimate is completely determined by the data segment under analysis, which is more robust to slowly varying non-stationary noise compared to off-source spectral estimation using, e.g., Welch's method. Details of the {\tt BayesLine} algorithm can be found in \cite{Littenberg:2014oda}, and an in-depth study describing its merits over using the Welch's method can be found in \cite{Chatziioannou:2019zvs}.

\section{Injections}
\label{sec:injections}
To understand the variation of the overlap $\langle \hh_\textrm{LI}| \hh_\textrm{BW} \rangle$ as a function of the system properties, we run {\tt LALInference} and {\tt BayesWave} on simulated \gw{} signals added to noise from the LIGO and Virgo detectors. A simulated signal that is added to noise is also called an ``injection". To perform injections, the instrument noise from the LIGO and Virgo detectors is combined with the simulated \cbc{} waveform to make the simulated observation data stream $\dd$. This is then analyzed by the {\tt BayesLine} algorithm which computes the median \psd{}, $S_n(f)$, that is used in the likelihood computations described in Equation \ref{eq:likelihoood}. $S_n(f)$ and $\dd$ are then fed into {\tt LALInference} and {\tt BayesWave} for analysis. This is exactly the same procedure as is used in LIGO and Virgo data offline \pe{} follow up analyses on actual \gw{} event detections.
We compute the overlap between $\hh_\textrm{LI}$ and $\hh_\textrm{BW}$ using Eq. \ref{eq:network_overlap}. 

We  apply the above analysis to two types of injections. The first type, ``GWTC-1 injections" are injections of signals from systems whose parameters are drawn from the posterior distribution samples of \gw{} events in GWTC-1. The purpose of these injections is to establish an expectation of the overlap for a each event in GWTC-1, which we then use to compare with the overlap on the actual event observation data.  The second type, referred to as ``Population injections", are injections of signals from systems with total mass $M_{T}$ in the range range $10 M_{\odot}$ to $120 M_{\odot}$. These help us establish typical trends in the overlap over a broad range of systems. 

The two types of injections yield complimentary inferences. GWTC-1 injections focus on \cbc{} systems specific to events in GWTC-1 and are designed to gauge the reconstruction performance of the catalog, whereas the population injections are designed to infer the trends in the overlap over the range of systems that we expect to detect in ground based detectors.

\subsection{Reconstruction of Detected Signals}
\label{sec:GWTC-1}
To test the reconstruction fidelity for real events, we design a set of $500$ injections, $50$ for each of the $10$ GWTC-1 events. The parameters of these injections are sampled from their measured {\tt IMRPhenomPv2} \cite{Hannam:2013oca} posterior probability density functions. We use the {\tt LALInference} posterior samples files available on the Gravitational Wave Open Science Center (GWOSC) \cite{Vallisneri_2015} and for each of the above injections, compute the ``offsource" overlap ($\cal{O}_\textrm{OFS}$ $= \langle \hh_\textrm{LI}| \hh_\textrm{BW} \rangle$). We then compare the distribution resulting from these $50$ $\cal{O}_\textrm{OFS}$ values with the ``onsource" overlap ($\cal{O}_\textrm{ONS}$ $= \langle \hh_\textrm{LI}| \hh_\textrm{BW} \rangle$) obtained from the data containing the real event. 

Since the parameters of these events are mostly consistent with nearly equal mass, spin-aligned systems with little to no evidence for precession, we expect the $\cal{O}_\textrm{ONS}$ value(s) to be no worse than the $\cal{O}_\textrm{OFS}$ overlaps(s).

We quantify this consistency using the p-value which we define as the fraction of $\cal{O}_\textrm{OFS}$ that are less than or equal to $\cal{O}_\textrm{ONS}$, i.e, $p \coloneqq P(\textrm{O}_\textrm{OFS} \leq \textrm{O}_\textrm{ONS})$. A smaller p-value indicates a smaller chance that the onsource reconstruction performance is consistent with what we expect. This could point to features in the onsource data that corrupt the reconstruction performance. These artifacts could be astrophysical or terrestrial in nature.

\subsection{Reconstruction Fidelity}
\label{sec:fidelity}
Past studies have shown that the agreement between burst and \cbc{} waveforms is most sensitive to the total mass $M_{T}$ of the \gw{} source and the \snr{} of the \gw{} signal, and monotically increases for both these quantities \cite{TheLIGOScientific:2016uux}. To systematically study the trends in the overlap as a function of these quantities, we analyze injections of a population of {\tt IMRPhenomPv2} waveforms using {\tt LALInference} and {\tt BayesWave}. We inject into noise from the second observing run of the LIGO detectors \cite{Vallisneri_2015}. We divide these ``population injections" into four different subsets based on their total mass $M_{T}$:  (i) $10\, M_{\odot} < M_{T} < 30\, M_{\odot}$, (ii) $30\, M_{\odot} < M_{T} < 60\, M_{\odot}$, (iii) $60\, M_{\odot} < M_{T} < 90\, M_{\odot}$, (iv) $90\, M_{\odot} < M_{T} < 120\, M_{\odot}$, the typical mass ranges we expect to observe in ground based detectors. The mass ratios, spins, orientations and sky locations were distributed uniformly. For each of the mass ranges, we created five population sets of \snr{s}: $10$, $20$, $30$, $60$, $90$. To strike a balance between compuational cost and number of sample points, we perform $50$ injections per \snr{} range per mass range, for a total of 1000 injections.

\section{Results}
\label{sec:results}


\subsection{GWTC-1 Injections}
\label{sec:gwtc1_results}

\begin{figure}[h]
    \centering
    \includegraphics[width=\columnwidth,clip=true]{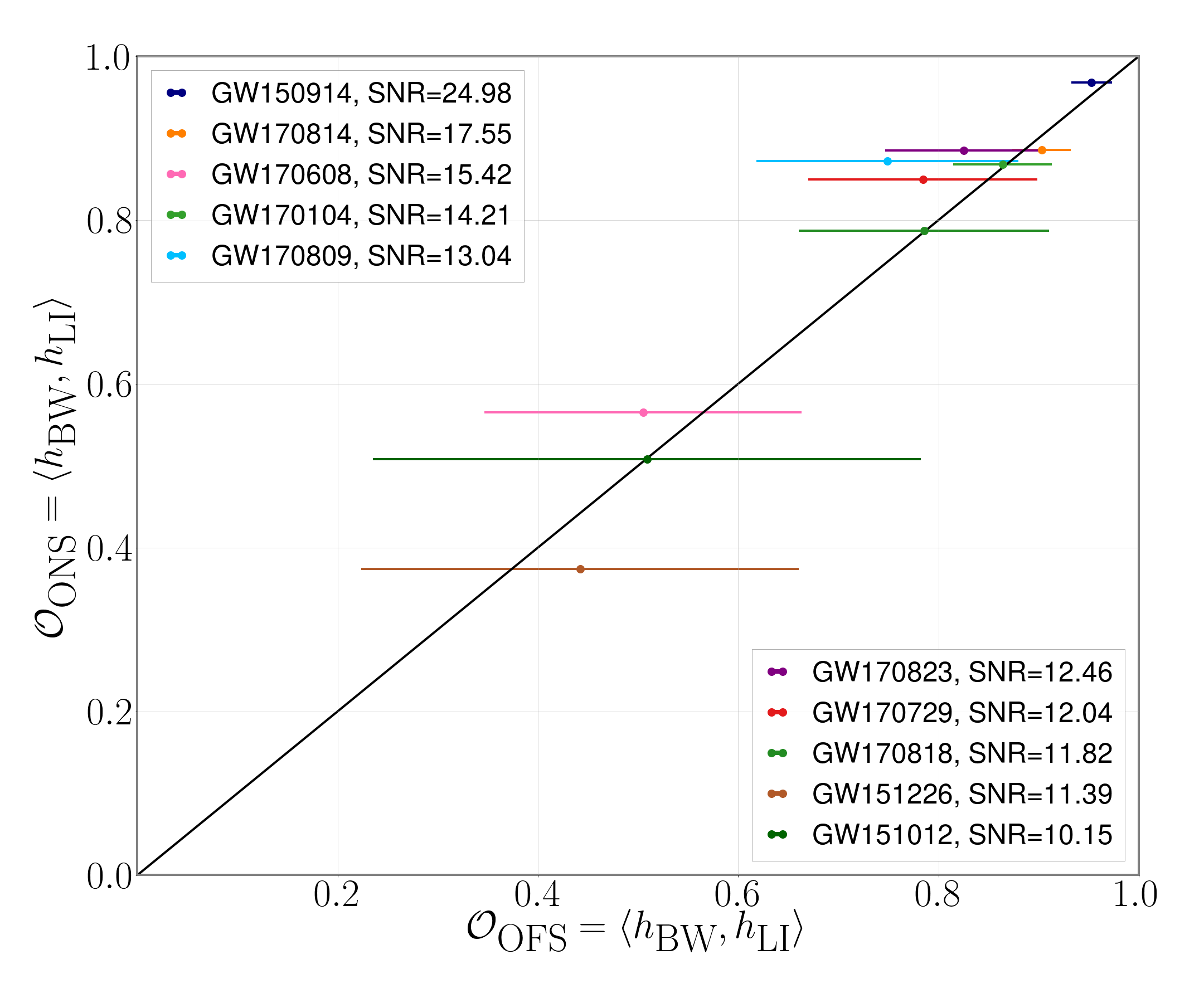}
    \caption{Onsource overlap (y axis) plotted against the median offsource overlap (x axis) for each of the GWTC-1 events. The horizontal error bars are the $90\%$ credible intervals in the overlap. The diagonal black line is $y=x$.} 
    \label{fig:onsource_v_offsource_gwtc1}
\end{figure}

We plot $\cal{O}_\textrm{ONS}$ as a function of $\cal{O}_\textrm{OFS}$ for each \bbh{} in GWTC-1 in Fig. \ref{fig:onsource_v_offsource_gwtc1}. Due to variations in the parameter posteriors and/or noise properties, the distribution of O$_\textrm{OFS}$ has a spread. The overlaid diagonal line here ($y=x$) represents the null hypothesis that the $\cal{O}_\textrm{OFS}$ and $\cal{O}_\textrm{ONS}$ are equal.  Dots represent the median and the horizontal error bars are the $90 \%$ credible intervals of the $\cal{O}_\textrm{OFS}$ distributions. From Fig. \ref{fig:onsource_v_offsource_gwtc1}, we find that all events are consistent with $y=x$ within $90 \%$ credibility.  The median of $\cal{O}_\textrm{OFS}$ decreases and its spread increases with decreasing \snr{} and $M_{T}$ of the event. For example, GW150914 with an \snr{} $\sim 24$ and $M_{T} \sim 65 M_{\odot}$ has a larger $\cal{O}_\textrm{OFS}$ median and a smaller spread compared to GW170823 which has a similar $M_{T}$ but has \snr{} $\sim 11$. Similarly, GW170729, with a $M_{T} \sim 84 M_{\odot}$ and \snr{} $\sim 10$ has a larger median value and a smaller spread compared to GW151012 with similar \snr{} but $M_{T} \sim 27 M_{\odot} $. 

\begin{figure}[h]
    \centering
    \includegraphics[width=\columnwidth,clip=true]{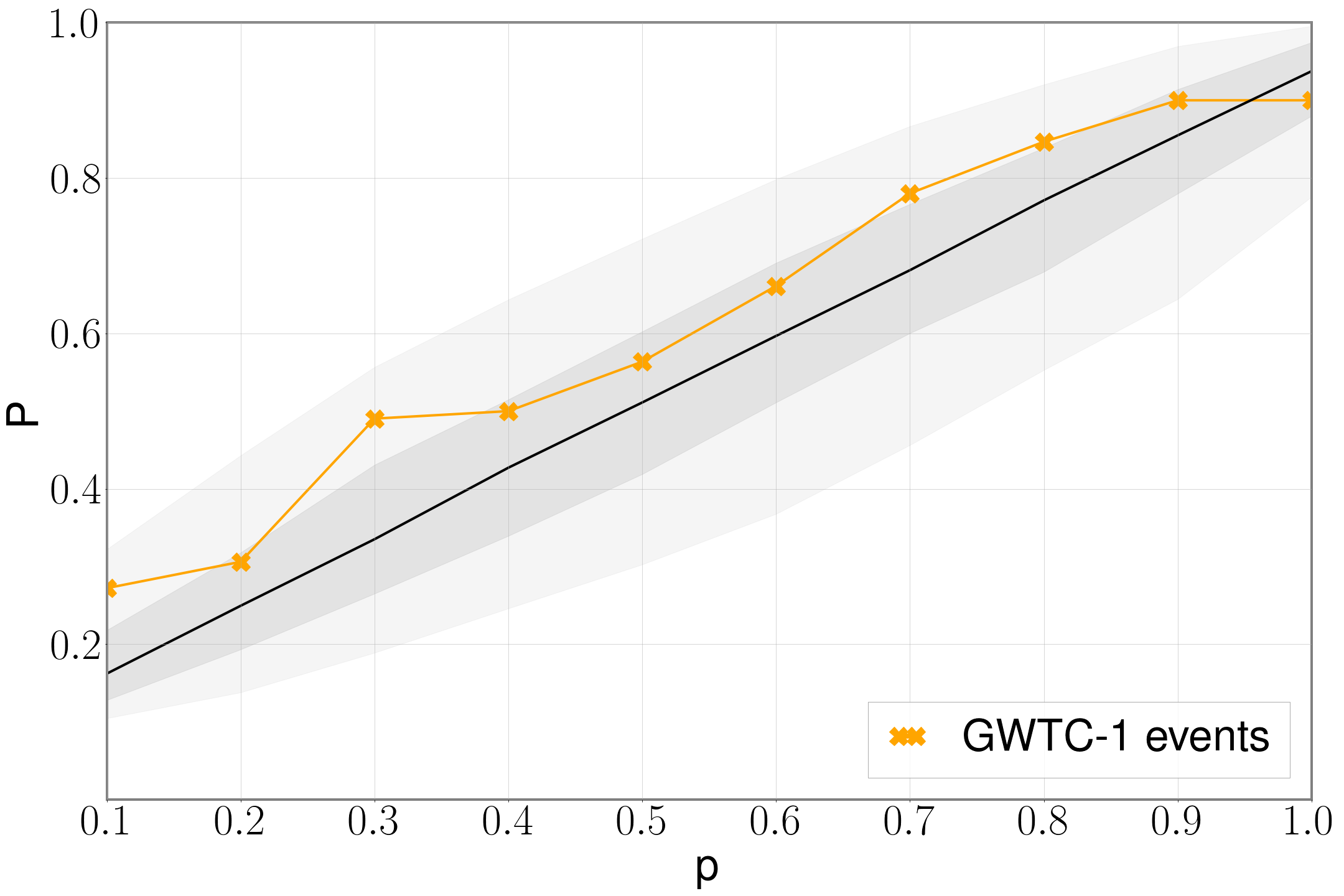}
    \caption{p-values (P) of GWTC-1 events (orange) plotted against the cumulative fraction events (p), along with the null hypothesis (black) and the 50\% and 90\% credible intervals (shaded regions).}
    \label{fig:p_value_gwtc1}
\end{figure}

We compute the p-values and record them in Table \ref{table:p-values}. We find that the p-values are broadly consistent with the null hypothesis that the onsource performance is no worse than the offsource performance. The lowest p-value that we compute is for GW151226 at $0.27$. Assuming the null hypothesis to be true for all events, we expect the p-values to be uniformly distributed between $0$ and $1$. We follow a procedure similar to \cite{LIGOScientific:2019fpa}, and use the Fisher's method \cite{10.2307/2681650} to compute the meta analysis p-value (p$_\textrm{meta}$) of the distribution of p-values. A p$_\textrm{meta}$ close to $1$ indicates higher evidence for the meta null hypothesis, and a p$_\textrm{meta}$ less than $0.05$ is considered low enough to reject the meta null hypothesis. We obtain a p$_\textrm{meta} = 0.95$ which indicates that there is no evidence for an aberrant behavior in the onsource reconstruction performance as compared to the offsource injections.

We also plot the p-values against the cumulative fraction of events in Fig. \ref{fig:p_value_gwtc1}. The black line represents the null hypothesis that the p-values are uniformly distributed, and the shaded bands represent the 50\% and 90\% credible intervals. The orange curve is consistent with the black line within the 90\% credible interval. Overall, this means that the agreement between burst and \cbc{} reconstructions for GWTC-1 events is statistically consistent with what we expect.

\begin{table}[]
\centering
  \begin{tabular}{l |c }
    \hline 
       Event&p-value\\ \hline 
       GW150914&$0.90$\\
       GW151012&$0.49$\\
       GW151226&$0.27$\\
       GW170104&$0.56$\\
       GW170608&$0.66$\\       
       GW170729&$0.78$\\
       GW170809&$0.90$\\
       GW170814&$0.30$\\
       GW170818&$0.50$\\
       GW170823&$0.84$
       
   \end{tabular}
   \caption{p-values of GWTC-1 events computed by comparing the onsource overlap of $\langle \hh_\textrm{LI}, \hh_\textrm{BW} \rangle$ versus the offsource distribution of the same}
   \label{table:p-values}
\end{table}

\subsection{Population Injections}
\label{population_results}

\begin{figure*}[]
    \centering
    \includegraphics[width=2\columnwidth,clip=true]{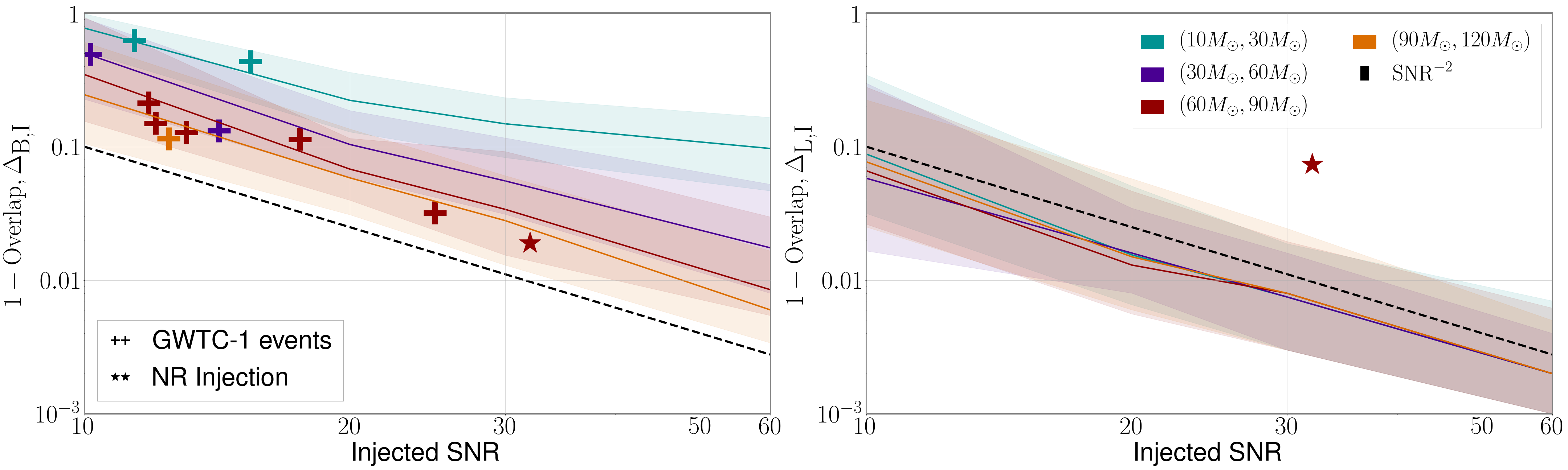}
    \caption{Medians (solid lines) and 90$\%$ uncertainty bands (shaded regions) of $\Delta_\textrm{B,I}$ (left) and $\Delta_\textrm{L,I}$(right) against the \snr{} ($\sqrt{\langle \hh_\textrm{INJ}, \hh_\textrm{INJ} \rangle}$) of the \cbc{} signal colored by the total mass ranges $M_{T}$. The black dashed-dot line represents the curve $1/\textrm{\snr{}}^2$ which is the expected variation of $\Delta_\textrm{B,I}$ and $\Delta_\textrm{L,I}$ for a signal. The overlaid ``$+$" markers indicate the onsource values of $\Delta_\textrm{B,L}$ plotted  against the \snr{} of the $\hh_\textrm{LI}$ waveforms for each of the GWTC-1 events, and are colored by the to $M_{T}$ of the $\hh_\textrm{LI}$ waveform. The overlaid ``$*$" markers indicate the values inferred from the \nr{} injection described in Section \ref{sec:disagreement}. } 
    \label{fig:bw_v_injection}
\end{figure*}
For each $M_{T}$ range and \snr{} pair, we compute of overlaps  between $\hh_\textrm{BW}$ and the injected \cbc{} signal, $\hh_\textrm{INJ}$ ($\cal{O}_\textrm{B,I}$), and the overlap between the $\hh_\textrm{LI}$ and $\hh_\textrm{INJ}$ ($\cal{O}_\textrm{L,I}$). As reconstruction performance improves, the overlaps become closer to $1$. For ease of visual interpretation, we define  $ \Delta = 1-\cal{O}$, where $\cal{O}$ is the overlap, and we use the same subscripts as for the overlap. $\Delta$ quantifies the disagreement between two waveforms. For each $M_{T}$ range, we obtain $\Delta$ distributions. We then plot the medians and 90$\%$ confidence intervals of these distributions as a function of the \snr{} in Fig. \ref{fig:bw_v_injection}. 

We see that at low \snr{s}, $\Delta_\textrm{B,I}$, where the subscripts ``B" and ``I" respectively represent {\tt BayesWave} and the injection, starts off high as {\tt BayesWave} is unable to recover the full signal. This is even more pronounced in systems with lower $M_{T}$ since the signal waveform is longer and the \snr{} is spread over a longer duration. We also see that at a particular \snr{}, $\Delta_\textrm{B,I}$ decreases with increasing $M_{T}$, since the signal waveform gets increasingly shorter and is more compactly represented with the wavelet model. $\Delta_\textrm{B,I}$ falls steadily as the \snr{} increases.  On the other hand, $\Delta_\textrm{L,I}$, where subscripts ``L" and ``I" represent {\tt LALInference} and the injection, is less than $0.2$, even for low \snr{s}, as {\tt LALInference} can reconstruct the \cbc{} signal morphology better than {\tt BayesWave} at lower \snr{}. This is expected, since {\tt LALInference} is using templates which predict the signal over the entire observing band. {\tt BayesWave} however, can only reconstruct high amplitude features in the data. $\Delta_\textrm{B,I}$  becomes smaller as $M_{T}$ and \snr{} increase, and {\tt BayesWave} is able to reconstruct more and more parts of the signal. Past studies have shown that we expect $\Delta_\textrm{B,I}$ and $\Delta_\textrm{L,I}$ to vary as $\propto 1/\textrm{\snr{}}^2$ \cite{Becsy:2016ofp}. We plot this curve and up to a constant scaling factor, we see that the slopes of the reconstructions follow this relationship to a large extent.

As an additional test of consistency, we overlay the values obtained from the onsource results of the GWTC-1 events in Fig. \ref{fig:bw_v_injection}. Specifically, we plot the $\Delta_\textrm{B,L}$, the complement of the overlap between $\hh_\textrm{BW}$ and $\hh_\textrm{LI}$ against the \snr{} of the $\hh_\textrm{LI}$ given by $\sqrt{\langle \hh_\textrm{LI}, \hh_\textrm{LI} \rangle}$, using $\hh_\textrm{LI}$ as a proxy for the true waveform. We justify this approximation by noting from Fig. \ref{fig:bw_v_injection} that $\Delta_\textrm{L,I}$ is less than $0.1$ which is an order a magnitude smaller than $\Delta_\textrm{B,I}$. The markers are colored according to the color scheme of the $M_{T}$ parameter as shown in Fig. \ref{fig:bw_v_injection}. We find that the $\Delta_\textrm{B,L}$ values fall within the bounds obtained from the population injections, which agrees with the inferences we drew in Section \ref{sec:gwtc1_results}.


\section{Detecting deviations}
\label{sec:disagreement}

\begin{figure*}[]
    \centering
    \includegraphics[width=2\columnwidth,clip=true]{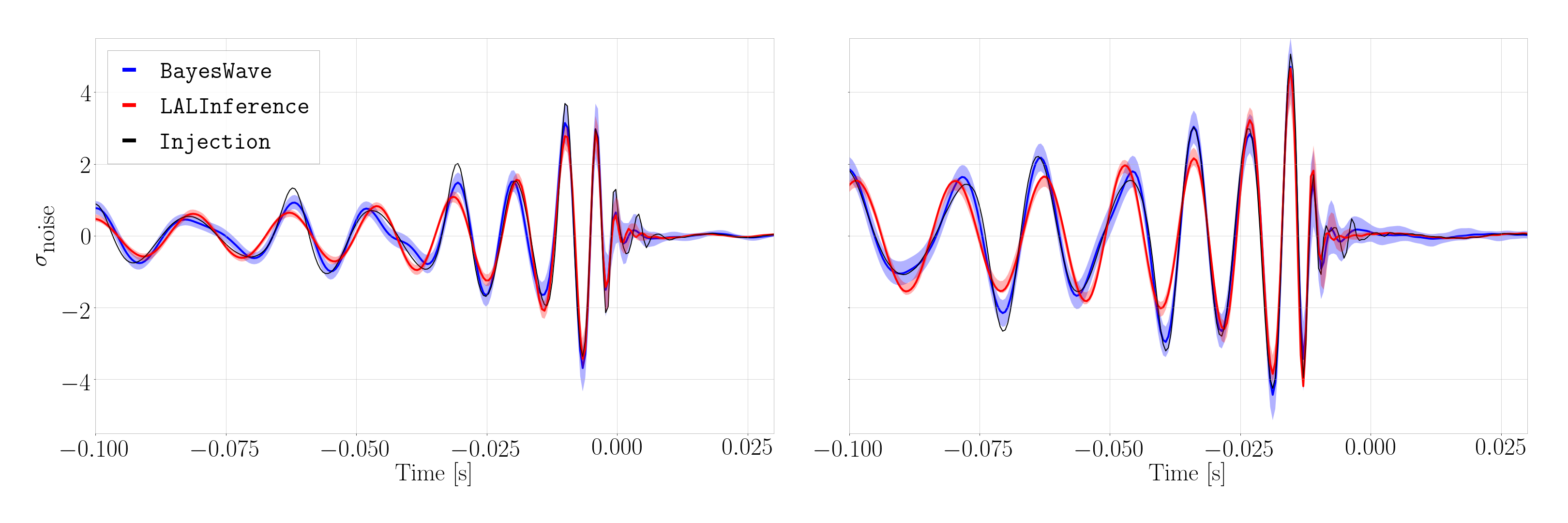}
   \caption{ {\tt LALInference}(red), {\tt BayesWave} (blue), and the injected waveform (black) for the injection analysis performed using the Georgia Tech \nr{} waveform {\tt GT0745}. The plotting conventions are similar to Fig. \ref{fig:gw150914_reconstruction}. Note the disagreement between $\hh_\textrm{LI}$ and $\hh_\textrm{BW}$, especially before $-0.4$ seconds.} 
    \label{fig:deviation}
\end{figure*}

Our analysis so far has been focused on the agreement between {\tt LALInference} and {\tt BayesWave} reconstructions. The results serve as a reference to check for consistency in future observations, and to identify outliers due to potential disagreements between reconstructions. These disagreements could arise for example due to \hom{s}, highly precessing orbits, deviations from \gr{} or noise. We demonstrate one such example of an injection of \bbh{} \gw{} signal containing \hom{s}.
In the \gr{}, \bbh{} signals are typically dominated by the $(l,|m|)=(2,2)$ spherical harmonic mode. This is true for most signals that are detectable by ground based detectors, and especially for binaries with comparable mass components observed face-on. {\tt IMRPhenomPv2} waveforms do not account for the presence of \hom{s}.

The relative power of \hom{s} to the dominant mode is most dependent on the mass ratio and the inclination angle \cite{Bustillo:2016gid}. To demonstrate this, we consider the case of \nr{}  simulation {\tt GT0745} from the Georgia Tech \nr{} catalog  \cite{Jani:2016wkt}. This system has a component mass ratio of $6:1$. We place the system in the ``edge-on" configuration where the angle between the line of sight and the normal vector to the plane of the orbit, known as the inclination angle, is $90^{\circ}$. A combination of unequal masses and edge-on inclination yields a high \hom{} content in the waveform. We also set the distance such that \snr{}$\sim 30$. 

We inject the waveform into a noise realization set equal to zero, and analyze the data stream using both {\tt LALInference} and {\tt BayesWave}. Since we assume that the noise is Gaussian, the expectation value of the noise stream $\textbf{n}$ over multiple noise realizations is $0$. Hence the performance the algorithms on zero noise data is the ``average" result over many noise realizations~\cite{Nissanke:2009kt}.

We compute the overlaps $\cal{O}_\textrm{B,I}$, $\cal{O}_\textrm{L,I}$ and plot all three waveforms in Fig. \ref{fig:deviation}. Inspecting the \snr{s} and overlaps in Table \ref{table:SNR_overlap}, we find that {\tt BayesWave} reconstructions the injection more faithfully than {\tt LALInference} that uses a model without \hom{}. This is also reflected in the fact that the former is able to recover a larger \snr{} compared to the latter. We plot $\Delta_\textrm{L,I}$ and $\Delta_\textrm{B,I}$ (red stars) in Fig. \ref{fig:bw_v_injection}. As one can see, the former is an outlier, while the latter falls within the confidence band based on expectations from simulated signals. In case of a real detection with potential unmodeled effects, it will not be possible to calculate $\Delta_\textrm{L,I}$ and $\Delta_\textrm{B,I}$ since we cannot know the true waveform. The quantity of interest is $\Delta_\textrm{B,L} = 1 - \cal{O}_\textrm{B,L}$. In this particular case, we compute the $\Delta_\textrm{B,L} = 0.06$ which lies outside the 90\% credible interval of the distribution of $\Delta_\textrm{B,L}$ obtained from the population injections.

We note that \cbc{} models that include \hom{s} exist \cite{London:2017bcn, Cotesta:2018fcv,Khan:2019kot}, and the above is only meant as an exercise to demonstrate how a disparity between {\tt LALInference} and {\tt BayesWave} would manifest itself.

\begin{table}[h]
\centering
  \begin{tabular}{l |c c c c c}
    \hline 
       IFO&LI \snr{}&BW \snr{}&Inj \snr{}&$\langle h_{\textrm{LI}}, h_{\textrm{INJ}} \rangle$&$\langle h_{\textrm{BW}}, h_{\textrm{INJ}} \rangle$\\ \hline 
       Hanford&$14$&$15$&$16$&$0.94$&$0.96$\\
       Livingston&$24$&$26$&$28$&$0.92$&$0.98$\\
       Network&$28$&$30$&$32$&$0.92$&$0.98$
   \end{tabular}
   \caption{\snr{s} and Overlaps for a {\tt LALInference} and {\tt BayesWave} analysis on an injection, Georgia Tech \nr{} waveform {\tt GT0745} with $M_{T}= 60 M_{\odot}$ and mass ratio, $q=6$, that includes \hom{s}. {\tt BayesWave} recovers a larger part of the waveform since {\tt LALInference} with {\tt IMRPhenomPv2} does not include \hom{s}.}
   \label{table:SNR_overlap}
\end{table}

\section{Conclusion and Discussion}
\label{sec:conclusion}
In this paper we systematically compared the reconstruction performance of a \cbc{} templated-based analysis, and a model-independent wavelet-based analysis for \bbh{} events.

We selected 50 random probability posterior parameter samples of each GWTC-1 \bbh{} event and injected them into LIGO and Virgo detector noise. We analyzed the injections using the {\tt LALInference} and {\tt BayesWave}, and checked consistency of the reconstructed waveforms by computing offsource overlaps. We computed the onsource overlap, and found that them to be consistent with the offsource overlaps within the $90 \%$ credible interval for all events. We also computed the p-values of the null hypothesis that the onsource overlap is no worse than the offsource overlap, and did not find any statistically significant evidence that suggests any deviation. The distribution of p-values obtained for all events yielded a meta analysis p-value of $0.95$ suggesting that the p-values are consistent with the meta null hypothesis that p-values are uniformly distributed. As a final step, we plotted the p-values in a p-p plot and found the distribution of p-values agrees with the null hypothesis that the p-values are uniformly distributed, within the $90 \%$ credible interval. All in all, this means that the GWTC-1 waveform reconstructions are consistent with expectations.

We also performed recovery on injections of a population of \bbh{} systems divided into bins of $M_T$ and \snr{}, and studied the overlap of the reconstructed {\tt LALInference} and {\tt BayesWave} with the true injected waveform and found that as expected, {\tt LALInference} is able to reconstruct the waveform more effectively than {\tt BayesWave} at all $M_T$ and \snr{}. The reconstruction performance increases with \snr{} for both the algorithms. Specifically, the $\Delta \sim 1/\textrm{\snr{}}^2$. $M_{T}$ does not have much effect on the reconstruction performance for {\tt LALInference} but the {\tt BayesWave} performance increases with increasing $M_{T}$. This was expected as higher total mass systems result inf high amplitude, short duration signals that {\tt BayesWave} is able to compactly represent with the wavelet model. We found that the onsource reconstruction performances of the GWTC-1 events are consistent with the trends inferred from the population injections.

Lastly, we demostrated an example of potential deviation from the above trends by injecting a waveform with strong \hom{s}, and studying its overlap with the {\tt LALInference} and {\tt BayesWave} reconstructions. We found that the {\tt LALInference} reconstruction, inferred using the {\tt IMRPhenomPv2} approximant, agrees less with the true waveform than the {\tt BayesWave} reconstruction, and stands as an outlier from the trends shown in Fig. \ref{fig:bw_v_injection}. The {\tt BayesWave} reconstruction is consistent with trends shown in Fig. \ref{fig:bw_v_injection}. This was expected since {\tt BayesWave} is agnostic to the physical aspects of the waveform morphology apart from speed of light propagation. 

We stress the importance of systematically characterizing the performance of the two algorithms on such systems that are challenging to model, for example where the $(l,|m| = (2,2))$ dominant spherical harmonic mode alone is insufficient to account for the signal morphology. With increasing sensitivity of the ground based detector network, any potential complex or mismodeled effects such as \hom{s}, high precession, or deviations from \gr{} could result in observable consequences and require more complete waveform models.

\section*{Acknowledgments}
\label{acknowledgments}
We would like to thank Christopher Berry and Benjamin Farr for lending us their software packages that we used in injection creation and post processing. We are also grateful to Jonah Kanner and Gregorio Carullo for their valuable comments on the manuscript.

This research has made use of data, software and/or web tools obtained from the Gravitational Wave Open Science Center (https://www.gw-openscience.org), a service of LIGO Laboratory, the LIGO Scientific Collaboration and the Virgo Collaboration. LIGO is funded by the U.S. National Science Foundation. Virgo is funded by the French Centre National de Recherche Scientifique (CNRS), the Italian Istituto Nazionale della Fisica Nucleare (INFN) and the Dutch Nikhef, with contributions by Polish and Hungarian institutes.
The authors are grateful for computational resources provided by the LIGO Laboratory and supported by National Science Foundation Grants PHY-0757058 and PHY-0823459.  

This research was done using resources provided by the Open Science Grid \cite{osg07, osg09}, which is supported by the National Science Foundation award 1148698, and the U.S. Department of Energy's Office of Science.

The GT authors gratefully acknowledge the NSF for financial support from Grants No. PHY 1806580, PHY 1809572, and TG-PHY120016.

The Flatiron Institute is supported by the Simons Foundation.

Parts of this research were conducted by the Australian Research Council Centre of Excellence for Gravitational Wave Discovery (OzGrav), through project number CE170100004.

NJC appreciates the support of NSF grant PHY1912053.




\bibliography{paper}
\end{document}